# Profiles for voltage-activated currents are multiphasic, not curvilinear

Per Nissen

Norwegian University of Life Sciences
Department of Ecology and Natural Research Management
P. O. Box 5003, NO-1432 Ås, Norway

per.nissen@nmbu.no

2016

# Abstract

Data for voltage-activation of a potassium channel (Matulef et al. *Proc Natl Acad Sci USA* 110: 17886-17891. 2013) were, as conventionally done, fitted by the authors by a Boltzmann function, i.e. by a curvilinear profile. Reanalysis of the data reveals however that this interpretation must be rejected in favor of a multiphasic profile, a series of straight lines separated by discontinuous transitions, quite often in the form of noncontiguities (jumps). In contrast to the generally very poor fits to the Boltzmann profiles, the fits to multiphasic profiles are very good. (For the four replicates, the average deviations from the Boltzmann curves were 10- to 100-fold larger than the deviations from the multiphasic profiles.) The difference in the median values was statistically highly significant, P<0.001 in most cases. For the mean values the deviations from the Boltzmann curve were 20-fold larger than the deviations from the multiphasic profile, and the difference in the median values was also highly significant. The curvilinear interpretation must be rejected also because of the uneven distribution of the points around the Boltzmann curves. The combined probability for the four replicates that this uneven distribution is due to chance is less than 0.002.

In addition to activation of ion channels, a wide variety of biological as well as non-biological processes and phenomena involving binding, pH, folding/unfolding and effect of chain length can be well represented by multiphasic profiles (Nissen 2015a,b. Posted on arXiv.org with Paper ID arXiv: 1511.06601 and 1512.02561).

# Introduction

In addition to multiphasic profiles for ion uptake in plants (Nissen 1971, 1974, 1991, 1996), such profiles have been recently (Nissen 2015a,b) reported for many other processes and phenomena. In the present paper, data (Matulef et al. 2013) for a potassium channel will be reanalyzed to statistically compare the fits to curvilinear profiles with the fits to multiphasic profiles.

In addition to the r values, slopes ± SE (or only slopes) are given on the plots. All slopes have been multiplied by 1000. The Runs test (Wald and Wolfowitz 1940) gives the probability for the uneven distribution of points around the curvilinear profile being due to chance.

# Reanalysis

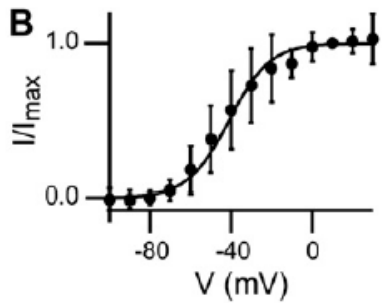

**Fig. 1.** Fig. S5B in Matulef et al. (2013). Voltage gating of the Y199-ester $K_vAP$ channel. Tail currents were recorded by stepping to -100 mV following the test pulse. The tail currents were normalized to the tail current observed with the +10-mV test pulse. The normalized tail currents were plotted against the test potential. The data were fitted with a Boltzmann function $I/I_{max} = 1/\{1 + \exp[-zF(V - V_{0.5})/RT]\}$ to give values of $V_{0.5} = -40.9 \pm 11.7$ mV and $z = 2.6 \pm 0.8$ (mean $\pm$ SD; $n = 4$). The solid line corresponds to a Boltzmann function plotted with the average values.

Original data kindly provided by Kimberly Matulef.

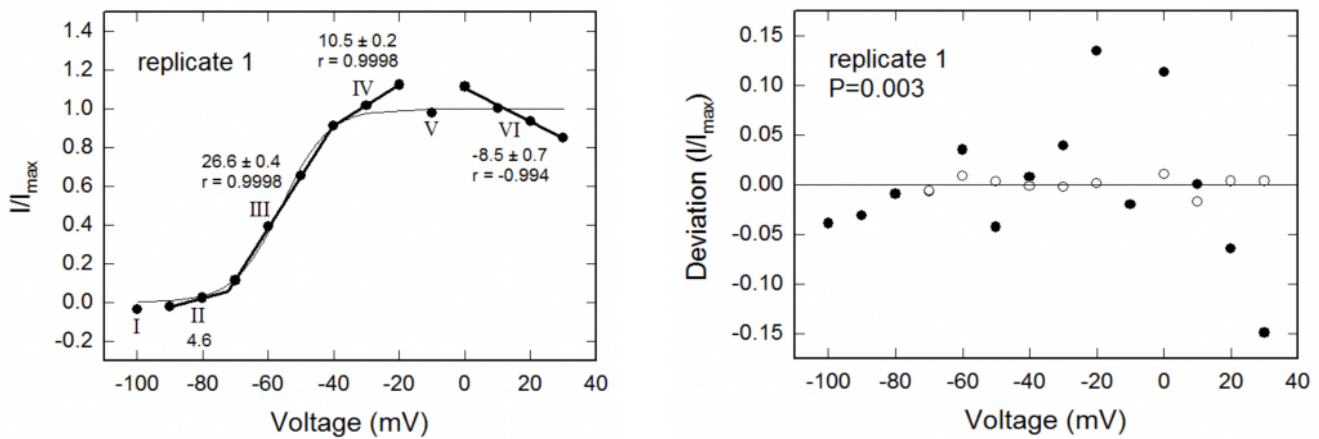

**Fig. 2** (left above). The profile can be well represented by 6 phases, with the transitions between -100 and -90, at -72.3 and -40, between -20 and -10, and between -10 and 0 mV. The data are insufficiently detailed in the range of phases I and V for resolution of the lines. High r values for lines III and IV. The Boltzmann curve gives in part very poor fits above -40 mV and cannot account for the marked decrease above 0 mV. The points are about evenly distributed around the curve. (The probability that the slightly uneven distribution is due to chance is 41.3%, by the Runs test.)

**Fig. 3** (right above). Plot of deviates for the data in Fig. 2. The deviations from the Boltzmann curve (filled circles) are on the whole much larger, 8.6-fold, than the deviations from the multiphasic profile (open circles). (The deviation at -70 mV is masked by the open circle.) The fits of the multiphasic profile are clearly better than the fits of the curvilinear profile, the difference in the median values is statistically highly significant (P = 0.003). If only the points for which there are deviations from the multiphasic profile are included, the deviations from the curvilinear profile are 10.3-fold larger than from the multiphasic profile, but the difference in fits is somewhat less significant (P = 0.011). It should be noted that the large deviations from the curve cannot somehow be ascribed to random errors, but result rather from the profile being multiphasic rather than curvilinear. Thus, the filled circles form a tetraphasic profile, with about straight lines from -100 to -60, from -50 to -20, from -20 to -10 and from 0 to 30 mV (the last two lines are about parallel).

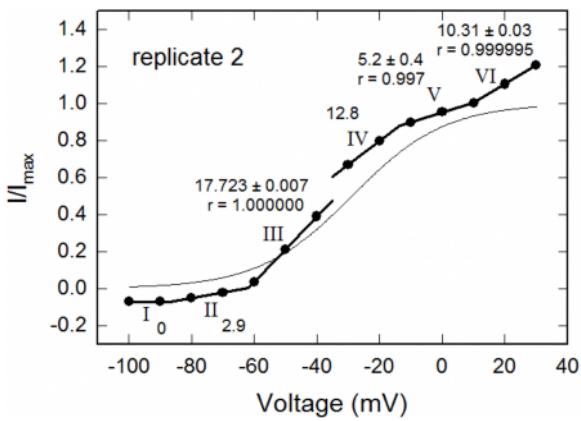 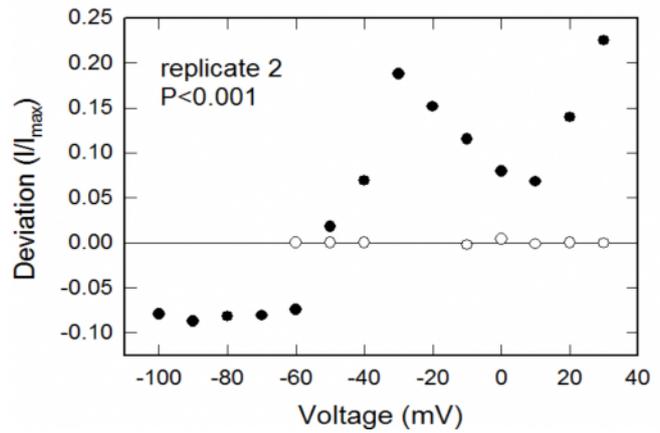

**Fig. 4** (left above). The profile can be well represented by 6 phases, with the transitions at -86.9 and -61.9, between -40 and -30 (jump), and at -13.4 and 10 mV. Exceedingly high r values for lines III and VI. Except at -50 mV, the Boltzmann curve gives very poor fits and cannot account for the marked increase at high voltages. The probability that the uneven distribution of points around this curve is due to chance is only 0.10%.

**Fig. 5** (right above). Plot of deviates for the data in Fig. 4. The deviations from the Boltzmann curve are very much larger, on the average 101-fold, than the deviations from the multiphasic profile (96-fold for the 8 points for which there are deviations from the multiphasic profile). The difference between the median values is in both cases highly significant (P<0.001). The deviations from the curvilinear profile form again a roughly tetraphasic profile.

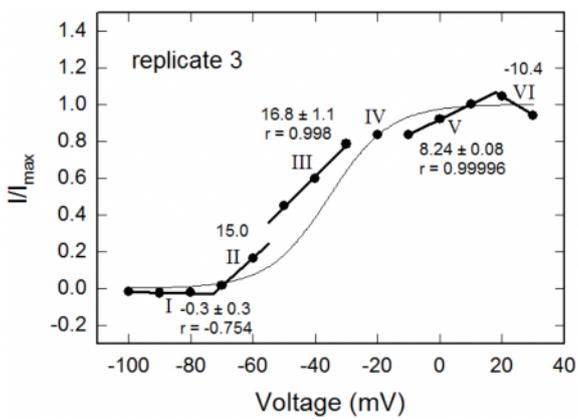 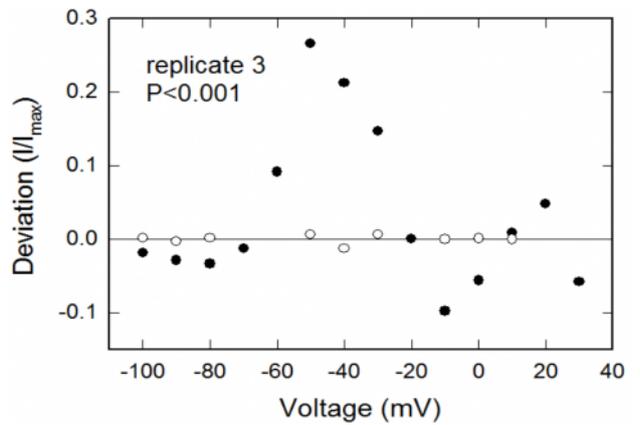

**Fig. 6** (left above). The profile can be well represented by 6 phases, with the transitions at -72.8, between -60 and -50 (jump), on both sides of -20, and at 18.0 mV. Insufficiently detailed data in the range of phase IV for resolution of the line. Very high r value for line V. Lines II and III are about parallel. At most voltages the Boltzmann curve gives very poor fits. Runs test: 7.8%.

**Fig. 7** (right above). Plot of deviates for the data in Fig. 6. The deviations from the Boltzmann curve are 49-fold (all points) or 34-fold (for 9 points) higher than the deviations from the multiphasic profile. The difference between the median values is in both cases highly significant (P<0.001). The deviations from the curvilinear profile form a tetra- or pentaphasic profile.

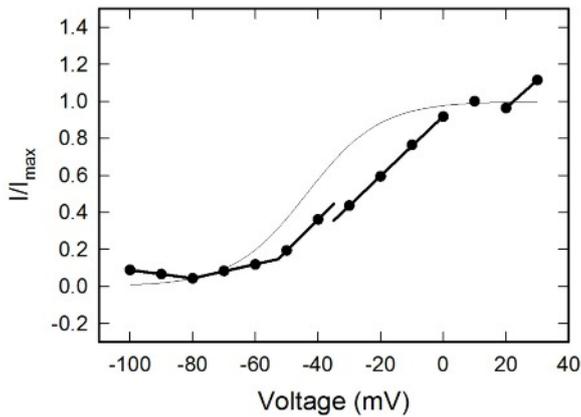 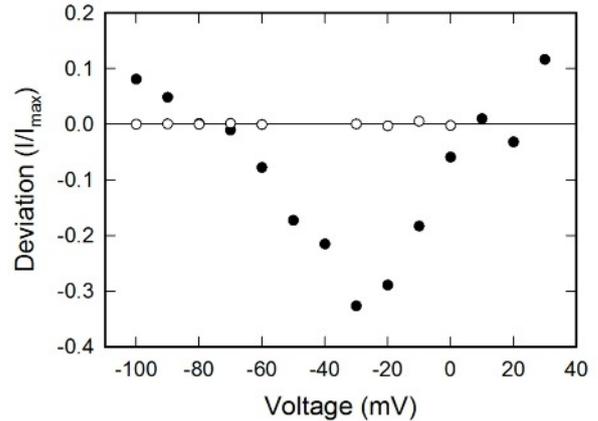

**Fig. 8** (left above). The profile can be well represented by 6 phases, with the transitions at -80 and -52.8, between -40 and -30 (jump), and on both sides of 10 mV. The data are insufficiently detailed in the range of phase V for resolution of the line. High absolute r values for lines I, II and IV. Lines III, IV and VI are about parallel. At most voltages the Boltzmann function gives very poor fits. The highly significant decrease in slope at low voltages (phase I) is also at variance with this function. Runs test: 11.9%.

**Fig. 9** (right above). Plot of deviates for the data in Fig. 8. The deviations from the Boltzmann curve are 74-fold (all points) or 76-fold (9 points) higher than the deviations from the multiphasic profile. The difference between the median values is in both cases highly significant (P<0.001). The deviations from the curvilinear profile form a tetra- or pentaphasic profile.

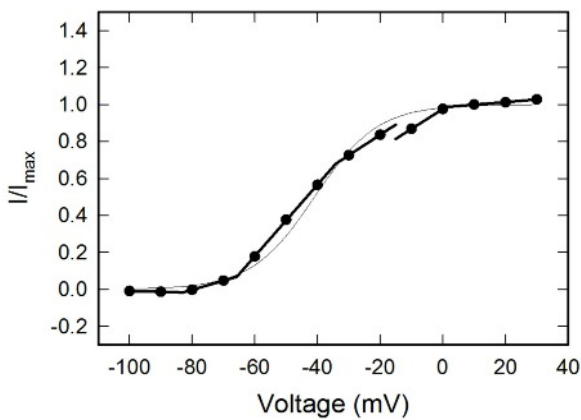 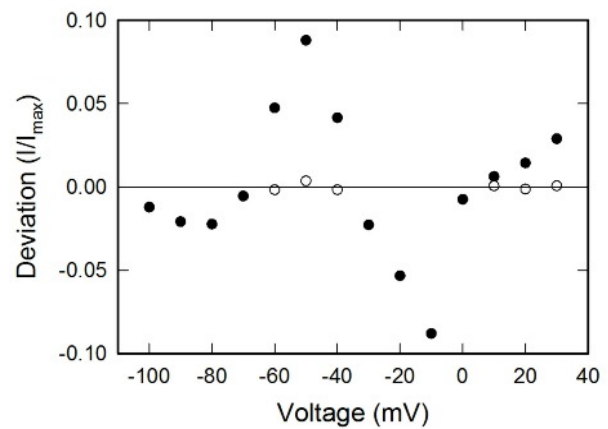

**Fig. 10** (left above). As also for each of the replicates, the profile can be well represented by 6 phases, with the transitions at -82.8, -65.8 and -34.2, between -20 and -10, and at 1.0 mV. High r value for line III. Lines IV and V are parallel.

**Fig. 11** (right above). Plot of deviates for the data in Fig. 10. The deviations from the Boltzmann curve are 20-fold (all points) or 23-fold (6 points) higher than the deviations from the multiphasic profile. The difference between the median values is in both cases highly significant, P<0.001 and P = 0.002, respectively. The deviations from the curvilinear profile form a tetra- or pentaphasic profile.

The profiles for the four replicates can all be well represented as hexaphasic, as can the profile for the mean. Otherwise, however, the profiles differ markedly from each other. Thus, there are in part large differences in the voltages at which the transitions occur. Furthermore, the decreasing slopes at high voltages for replicates 1 and 3 are not found in the profiles for the other replicates or for the mean. Also, only for replicate 4 is there a significant decrease in the slope at low voltages.

The representation of results by multiphasic profiles for means is, strictly speaking, meaningless when the profiles for the replicates differ as they do here and in many other sets of data. The means can give an indication, but only the profile for each replicate shows what is actually happening. Could it be that the conventional models, in some cases at least, somehow determine the overall shape of the profile, but at the same time the "fine structure" of the profile is multiphasic, with straight lines separated by discontinuous transitions (and with adjacent lines quite often being parallel)? In some systems, the variation between replicate experiments will average out to a smooth curve which itself represents the average physics and chemistry of the underlying system.

It seems that a curvilinear interpretation for the present data must be rejected not only from the comparison of the fits, but also from the finding that the points are unevenly distributed around the Boltzmann curves. The probability of this being due to chance is less than 0.002% for the four replicates (by a method for combining independent probabilities, Fisher 1954).

# Conclusion

From a comparison of fits it is clear, at high levels of significance, that the present data cannot be acceptably represented by curvilinear profiles. They are, instead, very well represented by multiphasic profiles. This conclusion holds also (Nissen 2015a) for other data (Schmidt et al. 2012, Devaraneni et al. 2013) for potassium channels as well as for a large body of other data (Nissen 2015a,b and in preparation).

**Acknowledgment** – I am grateful to Bob Eisenberg for his continued interest and encouragement.